\documentclass[aps,twocolumn,amsmath,amssymb,superscriptaddress,prb]{revtex4-2}
\usepackage{amssymb}
\usepackage{amsmath}
\usepackage{bm}
\usepackage{graphicx}
\usepackage{dcolumn}
\usepackage{color}
\usepackage{float}
\usepackage{epstopdf}

\newcommand{\be}{\begin{equation}}
\newcommand{\ee}{\end{equation}}
\newcommand{\bea}{\begin{eqnarray}}
\newcommand{\eea}{\end{eqnarray}}
\newcommand{\bs}{\begin{split}}
\newcommand{\bse}{\begin{subequations}}
\newcommand{\ese}{\end{subequations}}

\begin{document}
\title{Time reversal symmetry breaking and {\em s}-wave superconductivity in ${\rm CaPd_2Ge_2}$: A ${\mu}$SR study}

\author{V.~K.~Anand}
\altaffiliation{vivekkranand@gmail.com}
\affiliation{\mbox{Helmholtz-Zentrum Berlin f\"{u}r Materialien und Energie GmbH, Hahn-Meitner Platz 1, D-14109 Berlin, Germany}}
\affiliation{Department of Physics, University of Petroleum and Energy Studies, Dehradun, Uttarakhand, 248007, India}
\affiliation{Department of Mathematics and Physics, University of Stavanger, 4036 Stavanger, Norway}
\author{A. Bhattacharyya}
\altaffiliation{amitava.bhattacharyya@rkmvu.ac.in} 
\affiliation{Department of Physics, Ramakrishna Mission Vivekananda Educational and Research Institute, Belur Math, Howrah 711202, West Bengal, India}
\author{D.~T.~Adroja}
\affiliation{ISIS Facility, Rutherford Appleton Laboratory, Chilton, Didcot, Oxon, OX11 0QX, United Kingdom}
\affiliation {Highly Correlated Matter Research Group, Physics Department, University of Johannesburg, P.O. Box 524, Auckland Park 2006, South Africa}
\author{K. Panda}
\affiliation{Department of Physics, Ariel University, Ariel 40700, Israel}
\author{P.~K.~Biswas}
\affiliation{ISIS Facility, Rutherford Appleton Laboratory, Chilton, Didcot, Oxon, OX11 0QX, United Kingdom}
\author{A. D. Hillier}
\affiliation{ISIS Facility, Rutherford Appleton Laboratory, Chilton, Didcot, Oxon, OX11 0QX, United Kingdom}
\author{B. Lake}
\affiliation{\mbox{Helmholtz-Zentrum Berlin f\"{u}r Materialien und Energie GmbH, Hahn-Meitner Platz 1, D-14109 Berlin, Germany}}
 
\date{\today}
\begin{abstract}
${\rm CaPd_2Ge_2}$ which crystallizes in ${\rm ThCr_2Si_2}$-type body-centered tetragonal structure exhibits superconductivity below the critical temperature $T_{\rm c} = 1.69$~K\@.  We have investigated the superconducting gap structure and time reversal symmetry of the ground state in ${\rm CaPd_2Ge_2}$  by means of muon spin relaxation and rotation ($\mu$SR) measurements. Our analysis of $\mu$SR data collected in transverse magnetic field reveals BCS superconductivity with a single-band $s$-wave singlet pairing and an isotropic energy gap having the value $2\Delta(0)/k_{\rm B}T_{\rm c} = 3.50(1)$. Further, an increased relaxation rate in zero field $\mu$SR asymmetry spectra below $T_{\rm c} $ provides evidence for the presence of a spontaneous magnetic field in the superconducting state revealing that the time-reversal symmetry is broken in ${\rm CaPd_2Ge_2}$. \\ 
\end{abstract}

\maketitle

\section{\label{Intro} Introduction}

The 122-type compounds, owing to their simple ${\rm ThCr_2Si_2}$-type body-centered tetragonal (bct) structure, are of particular interests for the study of superconductivity. The FeAs-based 122 superconductors, such as ${\rm Ba(K)Fe_2As_2}$, are the well known examples where superconductivity is achieved by suppressing the spin-density wave ordering of Fe moments \cite{Johnston2010,Stewart2011,Chen2020,Baglo2022}. Recently some of us investigated superconductivity in ${\rm ThCr_2Si_2}$-type bct structured 122 compounds ${\rm CaPd_2As_2}$ and ${\rm CaPd_2Ge_2}$ \cite{Anand2013, Anand2014}. ${\rm CaPd_2As_2}$ exhibits  superconductivity below $T_{\rm c} = 1.27$~K \cite{Anand2013}. The measured and derived superconducting state parameters characterize ${\rm CaPd_2As_2}$ as a weakly coupled type-II $s$-wave superconductor in the dirty-limit \cite{Anand2013}. However, despite a very sharp superconducting transition in single crystal ${\rm CaPd_2As_2}$, the jump in the electronic specific heat at $T_{\rm c}$ reflects a value of $\Delta C_{\rm e}/\gamma_n T_{\rm c} = 1.14$ which is much smaller than the BCS expected value of 1.43. The reason for the reduced value of $\Delta C_{\rm e}/\gamma_n T_{\rm c}$ is not clear.

A reduced value of $\Delta C_{\rm e}/\gamma_n T_{\rm c} = 1.21$ is also seen in the case of ${\rm CaPd_2Ge_2}$ single crystal in which there is also a sharp and well pronounced jump in the electronic specific heat at $T_{\rm c} =1.69$~K \cite{Anand2014}. The superconducting state electronic specific heat data have been analyzed by the $\alpha$-model of BCS superconductivity \cite{Padamsee1973,Johnston2013}. For ${\rm CaPd_2Ge_2}$ the $\alpha$-model analysis yielded a value of $\alpha = \Delta(0)/k_{\rm B} T_{\rm c} = 1.62$ for $\Delta C_{\rm e}/\gamma_n T_{\rm c} = 1.21$ \cite{Anand2014}, which is lower than the BCS value $\alpha_{\rm BCS} = 1.764$.  For ${\rm CaPd_2As_2}$  a reduced value of $\alpha = 1.58$ was obtained for $\Delta C_{\rm e}/\gamma_n T_{\rm c} = 1.14$ \cite{Anand2013}. The reduced value of $\Delta C_{\rm e}/\gamma_n T_{\rm c}$  and hence $\alpha$ may be caused by an anisotropic superconducting energy gap, or due to the presence of multiple superconducting gaps \cite{Johnston2013}.

An ab initio calculation by the pseudopotential method and the generalised gradient approximation of density functional theory suggest that the major contribution to the density of states close to the Fermi level in ${\rm CaPd_2Ge_2}$ comes from Pd {\it d} and Ge {\it p} orbitals \cite{Karacaa2017}. Further, the estimate of electron-phonon interaction from the analysis of Eliashberg spectral function supports the conventional mechanism for superconductivity in ${\rm CaPd_2Ge_2}$. The electron and phonon couple through the vibration of Pd and Ge atoms. The vibration of Pd and Ge atoms modifies the tetrahedral bond angles in PdGe$_4$ tetrahedra in such a way that the Pd {\it d} and Ge {\it p} orbitals overlap \cite{Karacaa2017}. Such a change in tetrahedral bond angles in PdGe$_4$ and overlapping Pd {\it d} and Ge {\it p} orbitals is also seen in the Sr-analog ${\rm SrPd_2Ge_2}$ which exhibits superconductivity below 3 K \cite{Karacaa2016}

With an objective of shedding light on the nature of the superconducting gap structure in ${\rm CaPd_2Ge_2}$ we decided to examine the superconducting gap structure by the microscopic muon spin relaxation and rotation ($\mu$SR) measurements. Herein, we present the results of our $\mu$SR study on  ${\rm CaPd_2Ge_2}$.   The analysis of the $\mu$SR data suggests a single-gap isotropic s-wave superconductivity in ${\rm CaPd_2Ge_2}$. However, to our great surprise, the $\mu$SR  data reveals broken time-reversal symmetry in the superconducting state of  ${\rm CaPd_2Ge_2}$. A similar time-reversal symmetry broken superconducting state was inferred from $\mu$SR investigations on ${\rm Sc_5Co_4Si_{10}}$  \cite{Bhattacharyya2022} and other systems which also show an isotropic $s$-wave gap symmetry \cite{Singh2020, Barker2015La7Ir3, Arushi2022}. 

\section{\label{ExpDetails} Experimental Details}

A polycrystalline sample of ${\rm CaPd_2Ge_2}$ was prepared by the conventional solid state reaction method using the high purity starting materials [Ca-99.98\%, Pd-99.95\%, Ge-99.999\% from Alfa Aesar] at the Core Lab for Quantum Materials, Helmholtz-Zentrum Berlin (HZB). The Pd and Ge powders along with the Ca-pieces were pressed into pellet and sealed inside a quartz tube with pellet kept in an alumina crucible, which was then slowly heated to 800$^\circ$~C at a rate of 50$^\circ$ per hour and kept there for 30 hours, after which it was ground finely and again pelletized and sealed and subsequently heat treated at 900$^\circ$~C for 72 hours. This process of grinding, pelletizing and sealing was repeated again and heat treated at 900$^\circ$~C for another 72 hours. The quality of the sample synthesized this way was checked by powder x-ray diffraction (XRD) at room temperature using Cu K$_\alpha$ radiation with the Bruker-D8 laboratory-based x-ray diffractometer. The XRD data [see Appendix, Fig.~\ref{fig:XRD}] revealed the desired phase and confirmed the  ${\rm ThCr_2Si_2}$-type body-centered tetragonal structure of  ${\rm CaPd_2Ge_2}$. The lattice parameters obtained from the refinement $a = b = 4.3271(2)$~\AA\  and $c = 4.9823(7)$~\AA, and the $c$-axis position parameter $z_{Ge} = 0.376(7)$ are in good agreement with the respective values obtained for single crystal  ${\rm CaPd_2Ge_2}$ \cite{Anand2014} and with the literature values \cite{Venturini1989}.

The $\mu$SR measurements were carried out at the ISIS facility of the Rutherford Appleton Laboratory, Didcot, United Kingdom using the muon spectrometer MuSR which has 64 detectors for transverse and longitudinal applied field directions~\cite{Lee1999}. The ${\rm CaPd_2Ge_2}$  powder sample was mounted on a high purity Ag-plate (99.999\%). The use of Ag minimizes the contribution from the sample holder as the Ag gives only a non-relaxing signal. The powdered sample was mounted to Ag-plate by applying  the diluted GE varnish which was then covered with thin silver foils. As the muons are very sensitive to magnetic fields, correction coils were used to neutralize the stray fields to within 1~$\mu$T. $\mu$SR data were collected in both zero field (ZF) and applied transverse field (TF). The ZF-$\mu$SR measurements were made at several temperatures between 0.1~K to 2.5~K. The TF-$\mu$SR measurements were carried out between 0.1~K and 2.5~K in the presence of transverse fields of $H = $10, 20, 30, and 40~mT, which lie in between the lower critical field $H_\mathrm{c1} = 3.1$~mT  and upper critical field $H_\mathrm{c2} = 134$~mT \cite{Anand2014}. In the superconducting state with field cooled mode, muons probe the vortex lattice state.
The $\mu$SR spectra were analyzed with the program WiMDA~\cite{Pratt2000}.

\section{\label{Sec:muSR} Results and Discussion}

\subsection{\label{Sec:ZF} ZF $\mu$SR: time reversal symmetry state}

\begin{figure}
\includegraphics[width=\columnwidth]{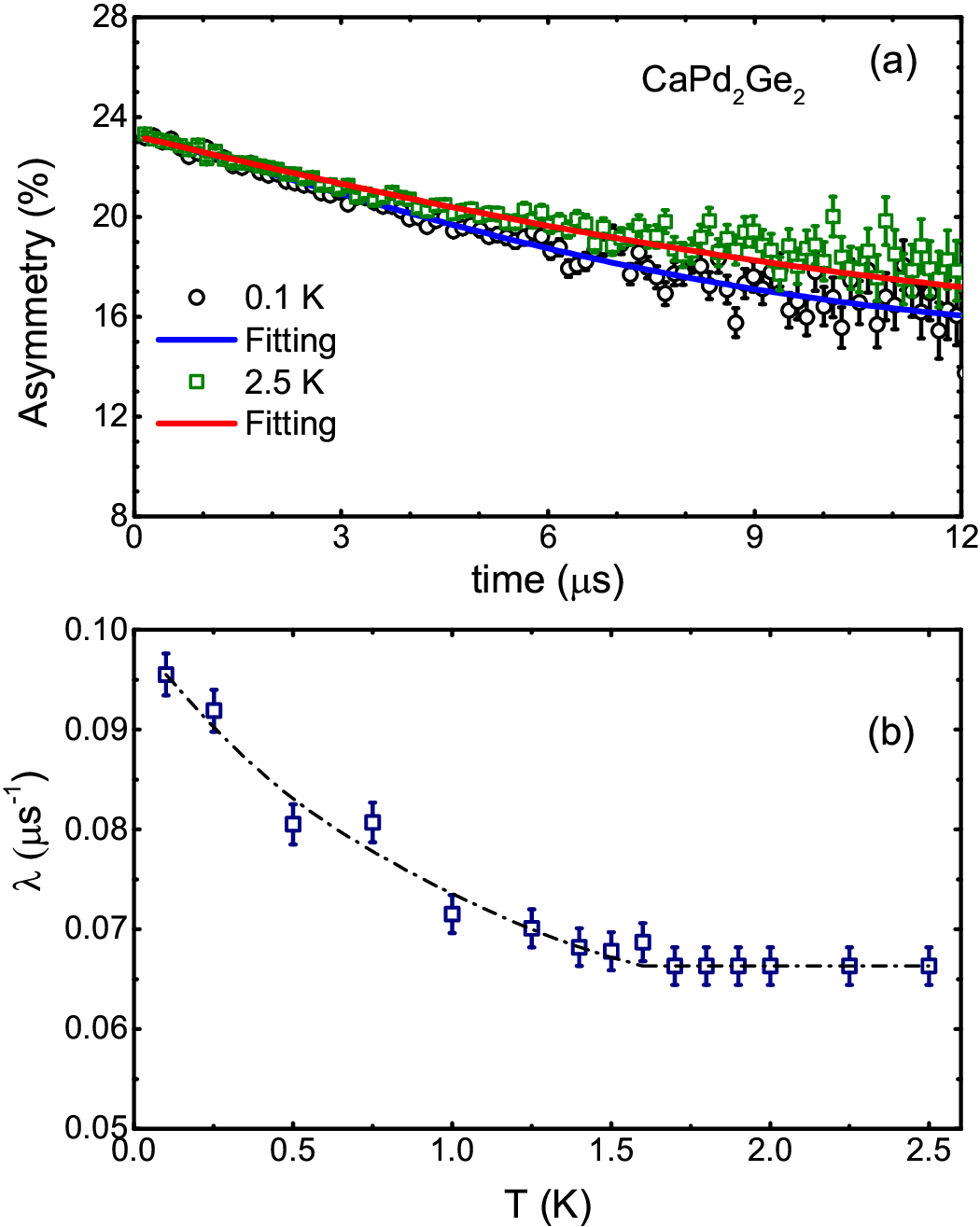}
\caption{(a) Zero field $\mu$SR time spectra for ${\rm CaPd_{2}Ge_2}$ collected at 0.1 K and 2.5~K\@. The solid curves represent the fit according to Eq.~(\ref{eq:MuSR_ZF}). (b) The temperature $T$ dependence of muon spin relaxation rate $\lambda_{\mathrm{ZF}}$ obtained from the fits $\mu$SR spectra collected at various $T$. The dashed navy blue line is the guide to the eye.}
\label{fig:MuSR_ZF}
\end{figure}

Figure \ref{fig:MuSR_ZF}(a) presents the representative $\mu$SR data collected in zero-field at 0.1 K (which is well below $T_{\rm c}$) and 2.5 K (which is well above  $T_{\rm c}$). The ZF $\mu$SR spectra allow us to discern the time reversal symmetry state of ${\rm CaPd_{2}Ge_2}$ by detecting the extremely weak magnetic field associated with the breaking of time-reversal symmetry.  
The time $t$ evolution of muon spin asymmetry spectra in zero field can be modelled by a damped Gaussian Kubo-Toyabe function \cite{Bhattacharyya2020Ce,HfIrSi},
\begin{equation}
\label{eq:MuSR_ZF}
 A_{\rm ZF}(t)=A_0\, G_{\rm KT}(t) \,{\rm e}^{-\lambda_{\rm ZF} t} + A_{\rm BG},
\end{equation}
where $A_0$ is  the initial muon asymmetry in zero field, 
\begin{equation}
\label{eq:KT}
 G_{\rm KT}(t)=\left[\frac{1}{3}+\frac{2}{3}\left(1-\sigma_{\rm KT}^2 t^2 \right){\rm e}^{ -\sigma_{\rm KT}^2 t^2/2}\right]
\end{equation}
is the Gaussian Kubo-Toyabe function, $\lambda_{\rm ZF}$ is the muon relaxation rate associated with the fluctuating fields due to electronic moments, and $A_{\rm BG}$ is the time-independent contribution from sample holder. In Eq.~(\ref{eq:KT}), the parameter $\sigma_{\rm KT}$ accounts for the Gaussian distribution of static fields associated with the nuclear moments.  The $\sigma_{\rm KT}$ is related to the local field distribution width as $H_{\mu} = \sigma_{\rm KT}/\gamma_{\mu}$ where $\gamma_{\mu}$ is the muon gyromagnetic ratio, $\gamma_{\mu}/2\pi$ = 135.53 MHz/T.

\begin{figure*}
\includegraphics[width=\textwidth]{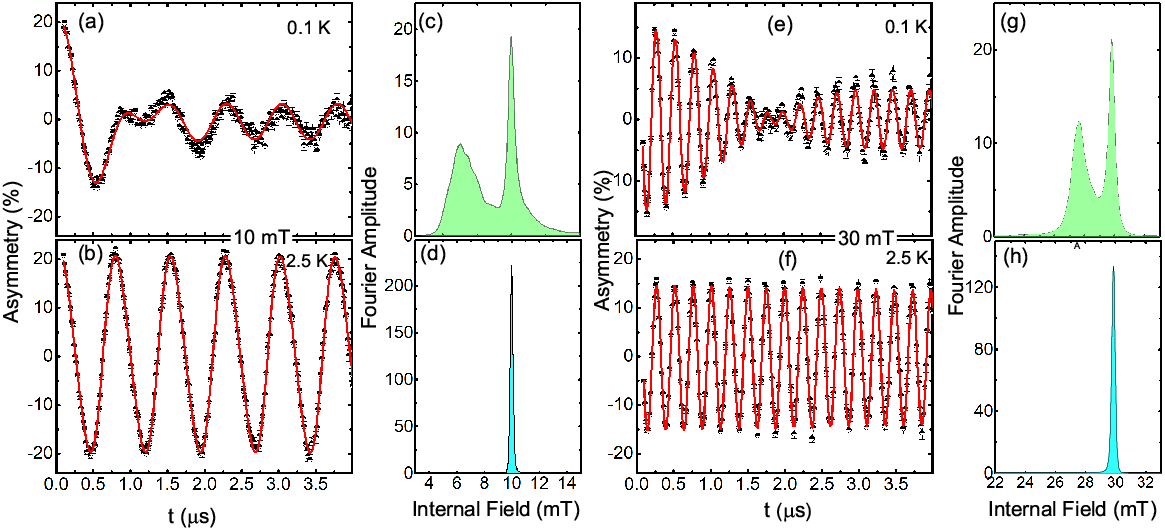} 
\caption{Transverse field $\mu$SR time spectra for ${\rm CaPd_{2}Ge_2}$ collected in the field-cooled state in an applied magnetic field of 10~mT at (a) 0.1~K and (b) 2.5~K, and that in a field of 30~mT at (e) 0.1 K and (f) 2.5~K along with their corresponding Fourier transformed maximum entropy spectra in (c), (d), (g) and (h), respectively. The solid red curve represents the fit according to Eq.~(\ref{eq:MuSR_TF}).}
\label{fig:MuSR_TF}
\end{figure*}

The representative fits of $\mu$SR spectra by the damped Gaussian Kubo-Toyabe function [Eq.~(\ref{eq:MuSR_ZF})] are shown by solid curves in Fig.~\ref{fig:MuSR_ZF} (a). The $\lambda_{\rm ZF}$ obtained from the fits of $\mu$SR spectra over the temperature range  0.1~K to 2.5~K are shown in Fig.~\ref{fig:MuSR_ZF} (b). The $T$ dependence of $\sigma_{\rm KT}$ obtained from the fitting is presented in Appendix [Fig.~\ref{Sigma}(a)]. At $T > T_{\rm c}$ the $\sigma_{\rm KT} \approx 0.0703~\mu$s$^{-1}$, and is nearly temperature independent over 0.1~K to 2.5~K. As can be seen from Fig.~\ref{fig:MuSR_ZF}(b), there is an increase in $\lambda_{\rm ZF}$ at  $ T <T_{\rm c}$. This kind of increase in $\lambda_{\rm ZF}$ indicates that the muons detect a spontaneous internal field while entering the superconducting state. Thus, the zero-field $\mu$SR measurements reveal that the time-reversal symmetry in the superconducting state is not preserved, and hence there is a time-reversal symmetry breaking in ${\rm CaPd_{2}Ge_2}$. The $\lambda_{\rm ZF}(T)$ seems to have an upward concavity which is not usual. Most of the superconductors with TRS broken superconducting state have been found to show a downward concavity in $\lambda_{\rm ZF}(T)$. The origin for this unusual upward concavity is not clear. An upward concavity in $\lambda_{\rm ZF}(T)$ was also observed in the case of fully gapped $s$-wave superconductor ScS with time-reversal symmetry broken superconducting state \cite{Arushi2022}.

Below $T_\mathrm{c}$, $\Delta\lambda_{\mathrm{ZF}}$ increases by 0.029(1)~$\mu$s$^{-1}$ corresponding to a characteristic magnetic field strength $\Delta\lambda_{\mathrm{ZF}}/\gamma_{\mu} = 0.034(2)$~mT, which provides clear evidence for the time reversal symmetry breaking in the superconducting state of CaPd$_{2}$Ge$_{2}$. Muon spin relaxation studies have provided evidence of TRS breaking in several superconducting materials like $\rm{Sr_2RuO_4}$ with a possible chiral $p$-wave symmetry \cite{Luke, Luke2000Sr2RuO4, Grinenko2019Sr2RuO4}, SrPtAs with a chiral $d$-wave symmetry \cite{SrPtAs}, LaNiC$_2$ with two nodeless gaps \cite{LaNiC2}, $A_{5}$Rh$_{6}$Sn$_{18}$ ($A =$~Y, $R$, or Sc) \cite{Sc5Rh6Sn18, Y5Rh6Sn18, Lu5Rh6Sn18}, ${\rm La_7(Ir, Rh, Pd)_3}$ \cite{Barker2015La7Ir3, Singh2020, Mayoh2021La7Pd3} and ${\rm Zr_3Ir}$ \cite{Shang2019Zr3Ir}  with a $s$-wave gap symmetry. For the present compound the spontaneous flux density due to superconductivity as estimated using the change in ZF-$\mu$SR relaxation rate above is 0.034(2)~mT. A similar estimate for Sr$_2$RuO$_4$ shows appearance of a characteristic field of 0.050~mT \cite{Luke}. The spontaneous field associated with TRS breaking in superconducting state of La$_7$Pd$_3$ is found to be 0.006~mT \cite{Mayoh2021La7Pd3} and that for Zr$_3$Ir to be 0.009~mT \cite{Shang2019Zr3Ir}.

Recently, we found evidence for TRS breaking in a centrosymmetric superconductor Sc$_5$Co$_4$Si$_{10}$ for which also an isotropic fully gapped s-wave symmetry was inferred from the $\mu$SR study \cite{Bhattacharyya2022}. The TRS breaking is usually associated with a non-unitary triplet pairing state or a mixed singlet-triplet state. However, a group theoretical analysis of Ginzburg-Landau’s free energy and symmetry allowed pairing states  using density functional theory does not support the presence of either non-unitary triplet pairing state or mixed singlet-triplet state in Sc$_5$Co$_4$Si$_{10}$ \cite{Bhattacharyya2022}. Accordingly, it was proposed that the Fermi surface topography of Sc$_5$Co$_4$Si$_{10}$ may allow TRS breaking by conventional electron-phonon mechanism \cite{Bhattacharyya2022}. We are under the impression that a similar physics associated with the Fermi surface topography could be held responsible for the observation of TRS breaking in the superconducting state of ${\rm CaPd_{2}Ge_2}$ with a fully gapped s-wave symmetry. 

Another mechanism which has been proposed for the TRS breaking in fully gap single-band BCS superconductors is based on loop super-current order applicable to the systems having complex crystal symmetry that allows the formation of microscopic super-current loops, such as in Re$_6X$ ($X =$ Zr, Hf and Ti) \cite{ghosh2020,Ghosh2021}. Ghosh et al. \cite{Ghosh2021} proposed that if the lattice unit cell of a superconducting material consists of more than two distinct symmetry-related sites, and the symmetry allows the order parameter to have different amplitude and phase at these symmetry sites, then it is possible to realize a superconducting ground state with spontaneous microscopic Josephson currents flowing between these symmetry sites of unit cell. These microscopic Josephson currents are termed as supercurrent loops, which, in turn produce a weak static magnetic field that can break the time-reversal symmetry and can be probed by $\mu$SR measurement which is extremely sensitive to the presence of even very weak magnetic field.  On the other hand, within the BCS formalism a TRS breaking in a multi-band superconductor can be associated with the development of a complex gap structure on account of the inter-band interactions \cite{wilson2013}.

\subsection{\label{Sec:TF} TF $\mu$SR: superconducting gap structure}

Figure~\ref{fig:MuSR_TF} presents the transverse field $\mu$SR asymmetry time spectra at 2.5~K (above $T_{\rm c}$) and 0.1~K (below $T_{\rm c}$) along with their Fourier transforms. The TF $\mu$SR data were collected in a field-cooled mode in the presence of applied magnetic fields of 10~mT and 30~mT.  It is clear from Figs.~\ref{fig:MuSR_TF}(a) and (e) that in the superconducting state ($T< T_{\rm c}$) the $\mu$SR spectra depolarize strongly on account of the inhomogeneous field distribution in the vortex state.

The TF $\mu$SR spectra could be fitted by a Gaussian oscillatory function and an oscillatory background, \cite{Adroja2021, BhattacharyyaThCoC2,comment}:
\begin{equation}
\label{eq:MuSR_TF}
\begin{split}
 A_{\rm TF} (t) &= A_1 \cos\left(\omega t + \phi \right) {\rm e}^{ -\sigma_{\rm TF}^2 t^2/2} \\
  & \hspace{1cm} + A_{\rm BG} \cos\left(\omega_{\rm BG} t + \phi \right),
\end{split}
\end{equation}
where $A_1$ is the initial asymmetry of the sample and $A_{\rm BG}$ that of the silver sample holder; $\omega = \gamma_{\mu}H_{\rm int}$, $H_{\rm int}$ being the internal field at muon site and $\omega_{BG} = \gamma_{\mu}H_{\rm int,BG}$; $\phi$ is the initial phase offset of the muon precession signal. $\sigma_{\rm TF}$ is the Gaussian relaxation rate that can be expressed as $ \sigma_{\rm TF}^2 = \sigma_{\rm sc}^2+\sigma_{\rm nm}^2$, where $\sigma_{\rm sc}$ accounts for the inhomogeneous field variation across the superconducting vortex lattice, and  $\sigma_{\rm nm}$ is the contribution due to the nuclear dipolar moments. The $\sigma_{\rm nm}$ was determined by fitting the spectra at $T>T_{\rm c}$ and kept fixed for $T < T_{\rm c}$, i.e.\ in the superconducting state, to obtain the value of $\sigma_{\rm sc}$ from the fit parameter $\sigma_{\rm TF}$.

The representative fits of the TF $\mu$SR spectra by the function discussed above in Eq.~(\ref{eq:MuSR_TF}) are shown by solid red curves in Figs.~\ref{fig:MuSR_TF}(a),(b),(e) and (f). The value of $\sigma_{\rm TF}$ is found to be much larger at $T<T_{\rm c}$ (superconducting state) than that at $T>T_{\rm c}$ (normal state) [see Appendix, Fig.~\ref{Sigma}(b)]. Further, the fitting of TF $\mu$SR spectra at 0.1~K and 10~mT by Eq.~(\ref{eq:MuSR_TF}) reveals that $A_1$ is 0.887 and $A_{\rm BG}$ is 0.113 of the total initial asymmetry, thus giving us an estimate of the lower bound of superconducting volume fraction of 88.7\%. This confirms the occurrence of bulk superconductivity in ${\rm CaPd_{2}Ge_2}$ as also inferred from the previous study \cite{Anand2014}.

The magnetic field probability distribution determined by the maximum entropy method is shown in Figs.~\ref{fig:MuSR_TF}(c), (d), (g) and (h) corresponding to the TF $\mu$SR spectra in Figs.~\ref{fig:MuSR_TF}(a), (b), (e) and (f), respectively. As can be seen from Figs.~\ref{fig:MuSR_TF}(d) and (h), at 2.5~K (in normal state) there is only one sharp peak at a value of $H_{\rm int}$ equal to the applied $H$. On the other hand, at 0.1~K (in the superconducting state), there is another broad peak at $H_{\rm int}$ lower than the applied $H$ in addition to the one at the applied $H$ [see Figs.~\ref{fig:MuSR_TF}(c) and (g)]. Such an appearance of an additional peak is a characteristic of type-II superconductivity. This inference of a type II behaviour in ${\rm CaPd_{2}Ge_2}$ is consistent with the estimated value of Ginzburg-Landau parameter ($\kappa_{\rm GL} = 6.3 > 1/\sqrt{2}$) \cite{Anand2014}.

\begin{figure*}
\includegraphics[width=\textwidth]{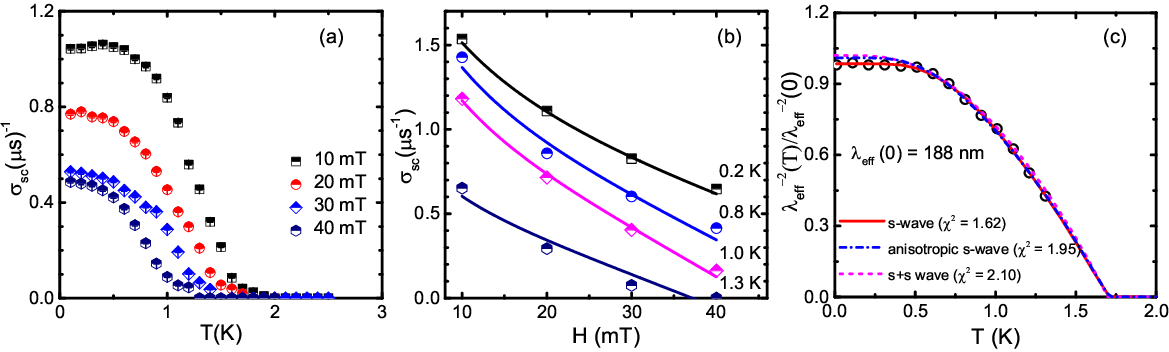} 
\caption{(a) Temperature $T$ dependence of the muon spin relaxation rate $\sigma_{\rm sc}$ for ${\rm CaPd_{2}Ge_2}$ collected in various applied transverse magnetic fields $H$  in field cooled state. (b) $H$ dependence of $\sigma_{\rm sc}$ at indicated temperatures. (c) $T$ dependence of inverse square of normalized penetration depth $\lambda_{\rm eff}$. The solid curve represents the fit for an isotropic single gap $s$-wave model according to Eq.~(\ref{eq:sigma_TF}).}
\label{TF}
\end{figure*}

The $T$ dependence of $\sigma_{\rm sc}$ extracted from the values of $\sigma_{\rm TF}$ that were obtained from the fit of the TF $\mu$SR spectra is shown in Fig.~\ref{TF}(a). The $H$ dependence of $\sigma_{\rm sc}$ is shown in Fig.~\ref{TF}(b). As the TF $\mu$SR spectra were collected at fields much lower than the upper critical field $H_\mathrm{c2}$, the Brandt \cite{Brandt1988,Brandt2003,ZrIrSi} relation 
\begin{equation}
\begin{split}
\sigma_\mathrm{sc} & = \frac{4.83 \times 10^{4}}{\lambda_{\rm eff}^{2}} (1-H_\mathrm{ext}/H_\mathrm{c2})  \\ 
& \hspace{1cm}\times [1+1.21\left(1-\sqrt{{(H_\mathrm{ext}/H_\mathrm{c2}})} \right)^{3}]
\end{split}
\label{eq:lambda}
\end{equation}
which holds good for $H/H_\mathrm{c2} \leq 0.25$, and for $\kappa_{\rm GL} \geq 5$, can be used to estimate the effective penetration depth $\lambda_{\rm eff}$. In this relation $\sigma_{\mathrm{sc}}$ is in the unit of $\mu s^{-1}$ and  $\lambda_{\rm eff}$ in nm.
For CaPd$_{2}$Ge$_{2}$  $\kappa_{\rm GL} = 6.3$ and $H_\mathrm{c2} = 134$~mT \cite{Anand2014}, therefore we can use the above Brandt relation [Eq.~(\ref{eq:lambda})] to estimate the  $\lambda_{\rm eff}$. The $T$ dependence of $\lambda_{\rm eff}$ obtained this way is  plotted as $\lambda_{\rm eff}^{-2}(T)/\lambda_{\rm eff}^{-2}(0)$ in Fig.~\ref{TF}(c). 

Apart from the information about the magnetic penetration depth, the $\sigma_{\rm sc}$ also provides information about the superfluid density, and size as well as the symmetry of the superconducting energy gap. In order to obtain information about the superconducting gap structure we analyzed the  $\lambda_{\rm eff}^{-2}(T)/\lambda_{\rm eff}^{-2}(0)$ by ~\cite{Prozorov, Bhattacharyya2021}
\bea
\frac{\sigma_{\rm sc}(T)}{\sigma_{\rm sc}(0)} &=& \frac{\lambda_{\rm eff}^{-2}(T,\Delta)}{\lambda_{\rm eff}^{-2}(0)} \nonumber\\
& = & 1 + \frac{1}{\pi} \int_{0}^{2\pi} \int_{\Delta(T, \varphi)}^{\infty}\frac{\partial f}{\partial E}\frac{E\,{\rm d}E\,{\rm d}\varphi}{\sqrt{E^2-\Delta^2(T, \varphi)}},
\label{eq:sigma_TF}
\eea
\noindent here $f$ is the Fermi function and $\varphi$ is the azimuthal angle in the direction of the Fermi surface. The Fermi function is given by $f=\left[1+\exp\left(E/k_{\mathrm{B}}T\right)\right]^{-1}$. The $T$ and $\varphi$ dependence of the order parameter $\Delta(T, \varphi)$ is given by $\Delta(T, \varphi) = \Delta(0) \delta(T/\it {T}_c)g(\varphi)$ \cite{Annett, Pang}. The angular dependence of the superconducting gap is contained in the function $g(\varphi)$. For $s$-wave BCS superconductivity with an isotropic gap, $g(\varphi)= 1$  \cite{Annett, Pang}. Further for the case of BCS superconductivity $\delta(T/T_c) =\tanh[(1.82){(1.018(T_c/T -1))}^{0.51}]$~\cite{UBe131}.

In order to determine the superconducting gap structure of ${\rm CaPd_2Ge_2}$ we analyzed the $T$ dependence of $\lambda_{\rm eff}^{-2}(T)/\lambda_{\rm eff}^{-2}(0)$ by Eq.~(\ref{eq:sigma_TF}) using three models: isotropic $s$-wave model, anisotropic $s$-wave model and $s+s$-wave model. For $s+s$-wave model, both the energy gaps were taken to be isotropic, and both the gaps were allowed to vary freely without any constraint. The fits for the three models are shown in Fig.~\ref{TF}(c) and the fitting parameters are listed in Table~\ref{Tabel}. It is evident from the fits in Fig.~\ref{TF}(c), as well as from the values of the quality of fit parameter $\chi^2$ in Table~\ref{Tabel}, that the single gap isotropic $s$-wave model describes the $\lambda_{\rm eff}(T)$ data better than anisotropic $s$-wave model and/or two-gap $s+s$-wave model. 

\begin{table}[b]
\caption{Fitting parameters obtained from the analysis of $\lambda_{\rm eff}^{-2}(T)/\lambda_{\rm eff}^{-2}(0)$ for CaPd$_{2}$Ge$_{2}$ according to Eq.~(\ref{eq:sigma_TF}) using three models: isotropic $s$-wave model, anisotropic $s$-wave model and $s+s$-wave model.}
\vspace{0.5cm}
\begin{tabular}{|c|c|c|c|}
\hline
{Model} &   {$\Delta_{i}(0)$(meV)} &{2$\Delta_{i}(0)$/k$_\mathrm{B}T_\mathrm{C}$}  & $\chi^{2}$  \\ \hline
isotropic $s$-wave & 0.252(3)  & 3.50(4) & 1.62(2) \\ 
\hline
anisotropic $s$-wave  & 0.311(4)       & 4.32(6) & 1.95(4) \\
\hline
two gap $s+s$-wave  & 0.281(2), 0.124(5)       & 3.90(1), 1.72(5) & 2.10(6)  \\
\hline
\end{tabular}
\label{Tabel}
\end{table}

The single gap isotropic $s$-wave model analysis, which describes the $T$ dependence of $\lambda_{\rm eff}$ the best, yielded an energy gap of $\Delta(0) = 0.25$~meV which corresponds to $2\Delta(0)/k_{\rm B}T_{\rm c} = 3.50(1)$. This value is quite close to the expected BCS weak coupling superconductor value of 3.53 but little higher than the value 3.24 obtained from the jump in specific heat at the superconducting transition \cite{Anand2014}. In addition, the $s$-wave analysis of $\lambda_{\rm eff}^{-2}(T)/\lambda_{\rm eff}^{-2}(0)$ provides an estimate of $\lambda_{\rm eff}(0) = 188(2)$~nm, which is consistent with the previous estimate of $\lambda_{\rm eff}(0) = 186(16)$~nm from the penetration depth measurement using tunnel diode resonator \cite{Anand2014}. In order to estimate the $\lambda_{\rm eff}$  according to Eqs.~(\ref{eq:lambda}) \& Eq.~(\ref{eq:sigma_TF}) we fitted the $\sigma_{\rm sc}$ values obtained from all the applied transverse fields. Further, following the approach detailed in Refs.~\cite{Hillier1997,Adroja2005, AnandLaIrSi3}, and assuming that approximately all the normal state carriers ($n_\mathrm{e}$) contribute to the superconductivity (i.e., $n_\mathrm{s} \approx$ n$_{e}$), we have estimated the superconducting carrier density $n_\mathrm{s}$. Using the relation $n_{\rm s} = m^*c^2/4\pi\lambda_{\rm eff}(0) e^2$, for $\lambda_{\rm eff}(0) = 188(2)$~nm, we obtained $n_\mathrm{s}$ = 1.20(1) $\times$ 10$^{27}$ carriers $m^{-3}$, where we used the value of effective mass $m^{*} = (1+ \lambda_{\rm e-ph}) m_{\rm e} \approx  1.51 \,m_\mathrm{e}$ with electron-phonon coupling constant $\lambda_{\rm e-ph} \approx 0.51 $ as estimated in Ref.~\cite{Anand2014} according to McMillan's relation \cite{McMillan1968}, and $m_{\rm e}$ being the free-electron mass.

Next we estimate the Fermi temperature $T_{\rm F}$ using the relation \cite{Kittel}
\begin{equation}
k_{\rm B} T_{\rm F} = \frac{\hbar^2}{2m^*} (3\pi^2 n_{\rm s})^{2/3},
\label{eq:TF}
\end{equation}
which, for $n_\mathrm{s}$ = 1.20(1) $\times$ 10$^{27}$ carriers $m^{-3}$, gives $T_{\rm F} = 3164$~K. This in turn gives the ratio $T_{\rm c}/T_{\rm F} \approx 0.0005$, and hence characterizes ${\rm CaPd_2Ge_2}$ as a conventional superconductor based on the Uemura plot \cite{Uemura1989,Uemera1991} that provides an empirical relation between $T_{\rm c}$ and $T_{\rm F}$ to classify a superconductor into the categories of conventional and unconventional superconductors. Uemura {\it et al}. \cite{Uemura1989,Uemera1991} plotted the values of ratio $T_{\rm c}/T_{\rm F}$ for many conventional and unconventional superconductors and suggested that an unconventional and exotic superconductivity is observed for those superconductors for which $0.01 \leq  T_{\rm c}/T_{\rm F} \leq 0.1$, whereas for a conventional superconductor $T_{\rm c}/T_{\rm F} \leq 0.001$. The value $T_{\rm c}/T_{\rm F} \approx 0.0005$ for ${\rm CaPd_2Ge_2}$ indeed falls in the range of conventional superconductor. Therefore the time reversal symmetry breaking in ${\rm CaPd_2Ge_2}$ is conjectured to have different physics than the one applicable to multiband or unconventional superconductors, or the systems with multiorbital character of states at the Fermi level.

\section{\label{Conclusion} Conclusions}

We have probed the superconducting gap structure and time reversal symmetry state of superconducting ${\rm CaPd_2Ge_2}$  through the muon spin relaxation and rotation measurements in both zero field and transverse magnetic field. The TF-$\mu$SR spectra were analyzed by a Gaussian oscillatory function, and the temperature and field dependences of muon-spin depolarization rate associated with the superconducting state $\sigma_{\rm sc}$ were obtained. Further, we obtained magnetic penetration depth from the $\sigma_{\rm sc}(T)$.  Information about the energy gap and pairing symmetry was obtained from the analysis of $\lambda_{\rm eff}(T)$. The $\lambda_{\rm eff}(T)$ is well described by single-band $s$-wave model indicating an isotropic superconducting gap structure. An energy gap of $2\Delta(0)/k_{\rm B}T_{\rm c} = 3.50(1)$ is obtained from the analysis. On the other hand, the analysis of ZF-$\mu$SR spectra revealed an increased relaxation rate below $T_{\rm c} $ on account of the spontaneous magnetic field associated with time reversal symmetry breaking in  ${\rm CaPd_2Ge_2}$. The observation of time-reversal symmetry breaking in  ${\rm CaPd_2Ge_2}$ is quite striking, and invites further experimental and theoretical investigations to understand the origin of this unconventional feature despite the conventional single-band isotopic $s$-wave singlet pairing symmetry of the superconducting order parameter in this compound.

\begin{figure} [b]
\includegraphics[width=\columnwidth]{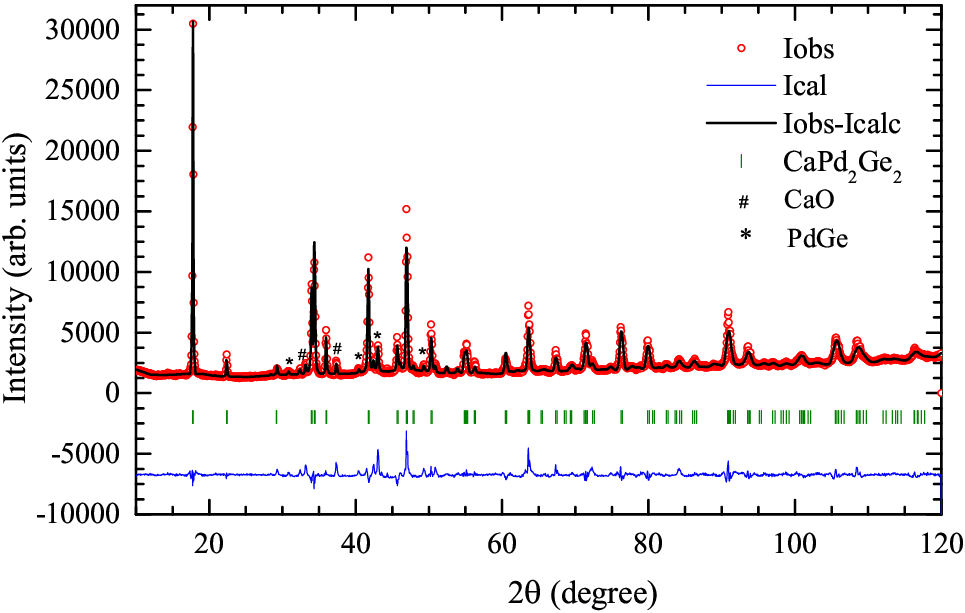}
\caption{\label{fig:XRD} Room temperature powder x-ray diffraction pattern (Cu K$_{\alpha}$ radiation) of CaPd$_2$Ge$_2$ along with the Rietveld refinement profile. The Rietveld refinement shown by the solid black curves represent major contribution from the bulk CaPd$_2$Ge$_2$. The Bragg peak positions for the CaPd$_2$Ge$_2$ are shown by the short vertical bars. The differences between the experimental and calculated intensities are shown by the lowermost blue curves. The impurity peaks marked with $\ast$ and \# represent the PdGe and CaO impurities, respectively.}
\end{figure}

\acknowledgements
 AB would like to thank Science \& Engineering Research Board for the CRG Research Grant (CRG/2020/000698). DTA would like to thank the Royal Society of London for the Newton Advanced fellowship funding between UK and China, and the International Exchange funding between UK and Japan. DTA also thanks EPSRC UK (Grant number EP/W00562X/1) for funding. We thank ISIS Facility for providing beam time RB1510100~\cite{DOI}.  

\section*{Appendix}

\begin{figure}
\includegraphics[width=\columnwidth]{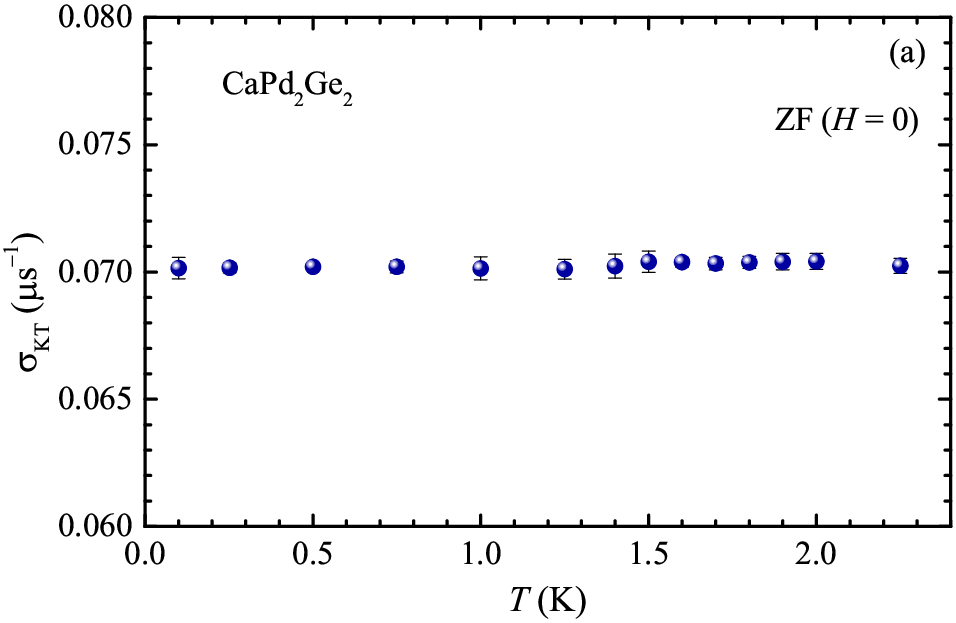}
\includegraphics[width=\columnwidth]{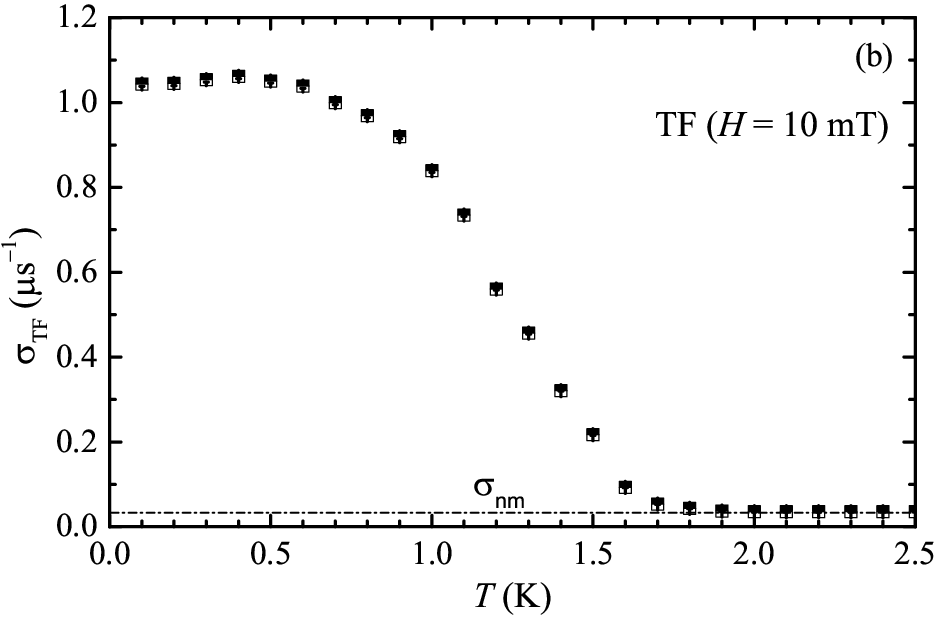}
\caption {(a)  The temperature $T$ dependence of KuboToyabe depolarization rate $\sigma_{\rm KT}$ obtained from the fitting of zero-field (ZF) $\mu$SR spectra of ${\rm CaPd_2Ge_2}$  by Eq.~(\ref{eq:MuSR_ZF}). (b) The $T$ dependence of Gaussian relaxation rate $\sigma_{\rm TF}$ obtained from the fitting of transverse-field (TF) $\mu$SR spectra measured in applied field $H=10$~mT by Eq.~(\ref{eq:MuSR_TF}). The $\sigma_{\rm nm}$ represents the contribution due to the nuclear dipolar moments.}
\label{Sigma}
\end{figure}

Figure~\ref{fig:XRD} shows the powder x-ray diffraction pattern of CaPd$_2$Ge$_2$ collected at room temperature along with the Rietveld refinement profile. The refinement confirmed the  ${\rm ThCr_2Si_2}$-type body-centered tetragonal structure of  ${\rm CaPd_2Ge_2}$. We also see few impurity peaks from PdGe and CaO, marked with  $\ast$ and \#, respectively, in Fig.~\ref{fig:XRD}. We estimate about 2\% CaO and 6\% PdGe, altogether about 8\% impurities in the present sample. The impurities are nonmagnetic in nature and therefore does not affect the results presented on the superconducting properties of CaPd$_2$Ge$_2$. The contribution from such spurious non-superconducting phase to $\mu$SR, if any, should be noticeable only in applied fields and not in zero-field. 

Figure~\ref{Sigma}(a) shows the temperature dependence of Kubo-Toyabe depolarization rate $\sigma_{\rm KT}$ obtained from the fitting of zero-field $\mu$SR spectra of ${\rm CaPd_2Ge_2}$ by Eq.~(\ref{eq:MuSR_ZF}). The $\sigma_{\rm KT}$ remains constant with a value of $ \approx 0.0703~\mu$s$^{-1}$. Figure~\ref{Sigma}(b) shows the temperature dependence of Gaussian relaxation rate $\sigma_{\rm TF}$ obtained from the fitting of transverse-field $\mu$SR spectra (for $H = 10$~mT) by Eq.~(\ref{eq:MuSR_TF}). The contribution due to the nuclear dipolar moments is also shown in Fig.~\ref{Sigma}(b). We find $\sigma_{\rm nm} \approx 0.0354~\mu$s$^{-1}$ at $T$ above $T_{\rm c}$.

\bibliographystyle{apsrev4-2}
\bibliography{refs}

\begin{thebibliography}{51}
\bibitem [{com()}]{comment}%
  \BibitemOpen
  \href@noop {} {}\bibinfo {note} {A comparison of Eqs. (1) and (3) reveal that
  the exponential multiplier of ZF fitting function is not required for the
  case of TF fitting function. This can be understood as: in TF measurements,
  the externally applied field is dominant and effectively suppresses the
  impact of any intrinsic or external fluctuating fields as well as Kubo-Toyabe
  contribution. Therefore, the exponential contribution, which accounts for the
  effects of fluctuating fields, which is very small of the order of $\mu$T
  becomes less relevant under the strong influence of the transverse field. For
  the present compound the fluctuating field of 0.034(2)~mT estimated from
  change in ZF-$\mu$SR relaxation rate is much smaller than the applied TFs of
  10~mT and 30~mT. For this reason, the multiplication by the exponential term
  can be neglected for TF fitting.}\BibitemShut {Stop}%
\end{thebibliography}
\end{document}